\documentclass[runningheads]{llncs}
\usepackage{cite}
\usepackage{amsmath,amssymb,amsfonts}
\usepackage{algorithmic}
\usepackage{textcomp}
\usepackage{xcolor}
\usepackage{svg}
\usepackage{xspace}
\usepackage[acronym]{glossaries}
\makeglossaries
\usepackage{multirow}
\usepackage{booktabs} 
\usepackage{url}
\usepackage{color}
\usepackage{multirow}
\usepackage{multicol}
\usepackage{enumitem}
\usepackage{array}
\usepackage[nolist,nohyperlinks]{acronym}
\usepackage{footnote}
\makesavenoteenv{tabular}
\usepackage{breqn}
\usepackage{fancyhdr}
\def\BibTeX{{\rm B\kern-.05em{\sc i\kern-.025em b}\kern-.08em
    T\kern-.1667em\lower.7ex\hbox{E}\kern-.125emX}}

\usepackage{graphicx}
\usepackage[acronym]{glossaries}
\acrodef{mIoU}{mean Intersection over Union}
\acrodef{DL}{deep learning}
\acrodef{ML}{machine learning}
\acrodef{DSC}{dice coefficient}
\usepackage[acronym]{glossaries}

\acrodef{mIoU}{mean Intersection over Union}
\acrodef{CADx}{Computer-Aided diagnosis}
\acrodef{CNN}{Convolutional Neural Networks}
\acrodef{FPS}{Frame Per Second}
\acrodef{WCE}{Wireless capsule endoscopy}
\acrodef{ASD}{Average Surface Distance}
\acrodef{AI}{Artificial Intelligence}
\acrodef{VCE}{Video capsule endoscopy}
\acrodef{DSC}{Dice Coefficient}
\acrodef{SOTA}{state-of-the-art}
\acrodef{HD}{Hausdorff distance}
\acrodef{MCC}{Matthews correlation coefficient}
\acrodef{GI}{gastrointestinal tract}
\acrodef{OAR}{Organs-at-risk}

\begin{document}
\title{Multi-Scale Fusion Methodologies for Head and Neck Tumor Segmentation}
%
%
\author{Abhishek Srivastava\inst{1}\orcidID{0000-0002-2418-6677} \and
Debesh Jha\inst{1}\orcidID{0000-0002-8078-6730} \and
Bulent Aydogan\inst{2}\orcidID{0000-0002-1322-7226} \and
Mohamed E. Abazeed\inst{3}\orcidID{0000-0002-6614-4440} \and
Ulas Bagci\inst{1}\orcidID{0000-0001-7379-6829}}
\authorrunning{Srivastava et al.}
%
\institute{Machine and Hybrid Intelligence Lab, Department of Radiology, Northwestern University, USA \\
\and {Department of Radiation Oncology, University of Chicago, Chicago, IL, USA}
\and{Department of Radiation Oncology, Northwestern University, Chicago, IL, USA}\\ }

\maketitle              
\begin{abstract}
Head and Neck (H\&N) organ-at-risk (OAR) and tumor segmentations are an essential component of radiation therapy planning. The varying anatomic locations and dimensions of H\&N nodal Gross Tumor Volumes (GTVn) and H\&N primary gross tumor volume (GTVp) are difficult to obtain due to lack of accurate and reliable delineation methods. The downstream effect of incorrect segmentation can result in unnecessary irradiation of normal organs. Towards a fully automated radiation therapy planning algorithm, we explore the efficacy of multi-scale fusion based deep learning architectures for accurately segmenting H\&N tumors from medical scans. Team Name: M\&H\_lab\_NU.
\end{abstract}
\section{Introduction}
Optimizations in radiation treatment plans for Head and Neck (H\&N) tumors have seen significant advancements in recent years. Quantitative imaging biomarkers obtained from medical scans have shown promise in modelling disease characteristics and treatment outcomes~\cite{lou2019image,aerts2014decoding}. A prerequisite to radiation therapy (RT) is an accurate delineation of (H\&N) tumors to obtain H\&N nodal gross tumor volumes (GTVn) and H\&N primary gross tumor volume (GTVp) from volumetric medical scans. Manual annotation of the region of interest requires significant content expertise and is both laborious and time-consuming, although being the gold standard. Instead, automated segmentation systems can swiftly provide segmentation maps of the region of interest and, consequently, improve patient care on a large scale. Since tumor size can vary, and the nature of the problem constitutes itself at varying scales, conventional deep learning algorithms provide only sub-optimal solutions for this problem. Recently, multi-scale fusion methodologies have shown great capacity in generating precise segmentation maps~\cite{wang2020deep,gu2021hrvit,srivastava2021gmsrf,srivastava2021msrf,srivastava2022efficient} when the object of interest exists in various different scales. Such methodologies have established their efficacy in the segmentation of 2-D medical images. The repeated fusion of multi-scale features generates diverse and robust features and allows a more generalizable model~\cite{srivastava2021gmsrf}, capable of modelling the varying size of the region of interest. We study the performance of such multi-scale fusion-based methodologies to obtain GTVn and GTVp from FluoroDeoxyGlucose (FDG)-Positron Emission Tomography (PET) and Computed Tomography (CT) scans. As participants in the HECKTOR 2022 challenge~\cite{andrearczyk2021overview,oreiller2022head}, we used the PET/CT images, GTVn masks, and GTVp masks released by the challenge organizers to train our two algorithms, named OARFocalFuseNet and 3D-MSF. We perform additional experiments with SwinUNETR~\cite{hatamizadeh2022swin}, to compare the efficiency of self-attention mechanisms by Transformers~\cite{dosovitskiy2020image,liu2021swin,khan2021transformers} with multi-scale fusion techniques. The organization of the rest of the paper is as follows. Section 2 provides a brief description of all the methods used for our experiments. Section 3 provides the experiment and implementation details. Section 4 discusses the results obtained by OARFocalFuseNet~\cite{srivastava2022efficient}, 3D-MSF~\cite{srivastava2022efficient}, and SwinUNETR~\cite{hatamizadeh2022swin}. Finally, we conclude our paper in Section 6.

\section{Method}
In this section, we discuss the three different deep-learning methodologies used in our experiments: OARFocalFuseNet~\cite{srivastava2022efficient}, 3D-MSF~\cite{srivastava2022efficient}, and SwinUNETR~\cite{hatamizadeh2022swin}.

\begin{figure*}[!t]
    \centering
    \includegraphics[width=0.8\textwidth]{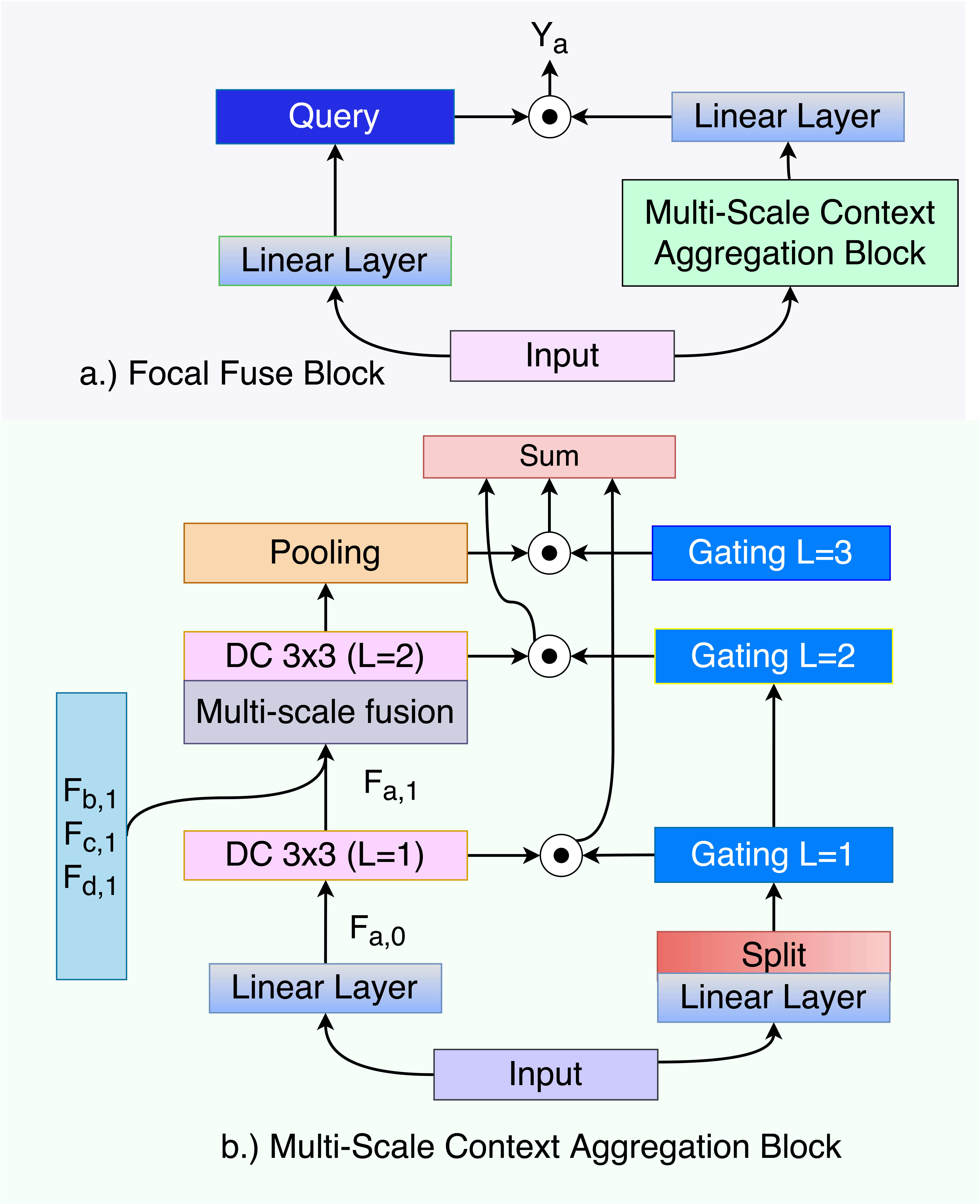}
    \caption{The OARFocalFuseNet components. {\bf a}) The Focal Fuse block, which aggregates multi-scale global-local context {\bf b}) The Multi-Scale Context Aggregation block, which gathers multi-scale features and performs depthwise convolutions to gather features with diverse context ranges and performs spatial and channel-wise gating to prune irrelevant features.}
    \label{fig:focal}
\end{figure*}

\subsection{Submission 1: OARFocalFuseNet}
\label{subsection:focalfuse}
Let $C$ and $P$ be the input CT and PET scan where $C \& P \, \in \, \mathbb{R}^{W\:\times\:H\:\times\:Z}$. Here, $W$, $H$, and $Z$ denote width, height, and length (number of slices), respectively. We concatenate $C$ and $P$ along the channel axis to form $X$ before feeding it into the encoder. The encoder blocks employ convolutional layers and pooling layers to extract features for a particular resolution scale and then downscale the resolution by a factor of 2. Let $[X_{1},X_{2},X_{3},X_{4}]$ be the sets of feature maps extracted by encoder blocks, each with a distinct resolution scale. Hereafter, a linear layer is used to transform the feature space $X_{a}$ into $F_{a,0}$, where $a$ denotes the resolution scale.

Multi-scale feature fusion is then performed by fusing multi-scale resolution features across all resolution streams (Figure~\ref{fig:focal}(b)). A combination of strided depth-wise convolution and pooling layers is used to downscale the spatial dimensions of features being transmitted from higher to lower-resolution streams. Similarly, a combination of strided depth-wise deconvolution and bicubic interpolation layers are used to upscale the spatial dimensions of features being transmitted from lower to higher-resolution streams.
\begin{multline}
    F_{a,l} = GeLU(DC_{3x3}(Conv_{1x1}(F_{a,l-1},F_{b,l-1},F_{c,l-1}, \\
    F_{d,l-1}))) \{b, c, d\} \neq a, \{a, b, c, d\} \in \{1,2,3,4\}.
\end{multline}
\label{eq:focal}
Here, $DC$ and $Conv$ represents a depth-wise convolutional layer and a standard convolutional layer, respectively. Additionally, $l$ denotes the multi-scale focal level, with the total number of focal levels being $N$. Moreover, a linear layer is utilized for pruning extraneous features (see equation~\ref{eq:gating}).
\begin{equation}
\label{eq:gating}
    G_{a,l} = Linear(F_{a,0}).
\end{equation}
where $G_{a}\, \in \, R^{W\:\times\:H\:\times\:Z\:\times\:N+1}$. The multi-scale focal modulator is calculated by adding the context information accumulated by each multi-scale focal layer (see equation~\ref{eq:aggregate} and Figure~\ref{fig:focal}(b)).
\begin{equation}
\label{eq:aggregate}
    F_{a} = \sum_{l=1}^{N+1} F_{a,l} \odot  G_{a,l}.
\end{equation}
Here, $\odot$ is an element-wise multiplication operator. The resultant focally modulated features for each scale are calculated as shown in Equation~\ref{eq:modeltoken} and Figure~\ref{fig:focal}(a).
\begin{equation}
\label{eq:modeltoken}
    MOD_{a} = \sum_{a=1}^{4} F_{a} \odot  Linear(I_{a}).
\end{equation}

\begin{figure*}[!t]
    \centering
    \includegraphics[width=0.8\textwidth]{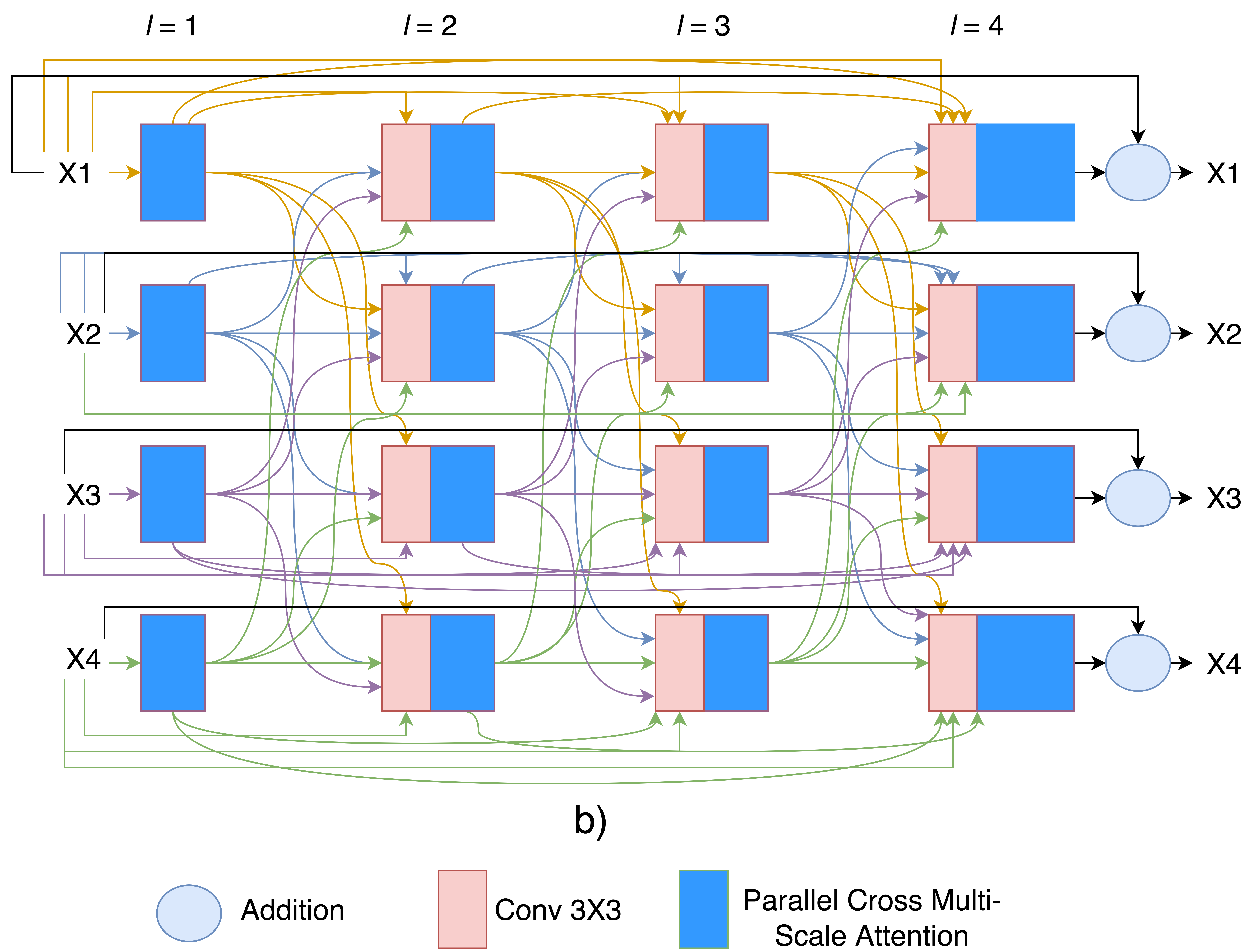}
    \caption{The multi-scale fusion module used in 3D-MSF. The colored lines illustrate feature fusion at multiple cross-scales. Here, each layer $l$ receives features from all preceding layers within the same resolution stream and the previous layer of all other resolution streams.}
    \label{fig:local}
\end{figure*}

\subsection{Submission 2: 3D Multi-scale Fusion Network}
The 3D Multi-scale Fusion Network (3D-MSF) uses densely connected blocks to perform multi-scale feature fusion. Initially, an encoder identical to the one used by OARFocalFuseNet is used for feature extraction. Each set of feature maps with a distinct resolution has its own resolution stream, which comprises a densely connected block. In this block, each layer receives inputs from all preceding layers in the same resolution scale and the previous layer from all other resolution scales (see Equation~\ref{eq:1} and Figure~\ref{fig:local}).
\begin{equation}
\begin{split}
    X_{a,l} = DepthConv_{3\times3}(X_{a,0} \oplus\cdots X_{a,l-1}  \oplus X_{b,l-1} 
    \oplus X_{c,l-1} \\\oplus X_{d,l-1}) , \{b, c, d\} \neq a, \{a,b,c,d\} \in \{1,2,3,4\},
    \end{split}
    \label{eq:1}
\end{equation}
where $l$ denotes the layer inside the dense blocks.

\subsection{Submission 3: SwinUNETR}
SwinUNETR follows the same architectural design as a standard UNet~\cite{ronneberger2015u}. SwinUNETR consists of an encoder, bottleneck, decoder, and skip connections. The basic unit of SwinUNETR is the Swin Transformer block. The input is first to split into non-overlapping patches before being projected to another feature space using a linear layer. Subsequently, these patches pass through the patch merging and Swin Transformer blocks to extract features. Then the decoder uses Swin Transformer blocks and patch-expanding layers to upscale the features obtained from the encoder. Additionally, the features obtained from the decoder blocks are fused with the corresponding encoder features via skip-connections. Lastly, the last patch expanding layer is used to perform 4× up-sampling to restore the resolution of the feature maps to the input resolution (WxHxZ). The Swin Transformer block combines the window-based multi-head self-attention (W-MSA)~\cite{liu2021swin} module and the shifted window-based multi-head self-attention (SW-MSA)~\cite{liu2021swin} module in succession before performing the self-attention operation.

\section{Experiments}
\subsection{Data Pre-processing and Data Augmentation}
We use the maximum of the CT/PET origin and the minimum of the CT/PET size to crop the input CT and PET volumes. Hereafter, they both are re-sampled and set to have the same origin, direction and size. Next, we perform the standard practice of clipping all CT values greater than 300 and less than -300 before performing min-max scaling. The PET volumes are also normalized using min-max scaling before being concatenated with CT volumes along the channel axis. Our data augmentation scheme involves random cropping, random Affine transformation, random 3D elastic transformation, and random Gaussian noise addition.
\subsection{Training Details}
We reserve 80\% of the data for training and 20\% of the data for validation. Each model is trained for 10,000 iterations and after every 500 iterations, performance on the validation set is evaluated. The best-performing model on the validation set is used for generating the final prediction masks. Adam-optimizer is used along with a cyclic learning rate scheduler with a base learning rate of 0.0005 and a maximum learning rate of 0.003. A batch size of 1 is used and the base filters for OARFocalFuseNet, 3D-MSF, and SwinUNETR are 16,16, and 48 respectively. We use an equally weighted combination of binary cross-entropy loss (see Equation~\ref{eq:bce}) and dice loss (see Equation~\ref{eq:dcs}). Here, $y$ is the ground truth value and $\hat{y}$ is the predicted value.
\begin{equation}
    \mathcal{L}_{BCE} = (y-1) \log (1 - \hat{y}) - y \log \hat{y}, 
   \label{eq:bce}
\end{equation}
\begin{equation}
\mathcal{L}_{DSC} = 1 - \frac{2y\hat{y}+1}{y+\hat{y}+1}.
\label{eq:dcs}
\end{equation}

\section{Results and Discussion}
In this section, we present the comparisons of the selected baselines on our validation set. We report the quantitative evaluation in Table~\ref{tab:result1}. Here, aggregated \ac{DSC} is used as the metric for evaluating our results. From Table~\ref{tab:result1}, we can observe that 3D-MSF obtains the highest aggregated \ac{DSC}, highest class-wise \ac{DSC} on GTVp and GTVn. Meanwhile, OARFocalFuseNet is able to outperform SwinUNETR in terms of aggregated \ac{DSC} and \ac{DSC} obtained on GTVn. Thus, multi-scale fusion methodologies are able to report significant performance gains over other \ac{SOTA} methods and can be developed further for tumor segmentation in H\&N CT/PET scans.
\begin{table}[!h]
\centering
\footnotesize
\caption{Result comparison on Hecktor 2022 Head and Neck Tumor segmentation challenge over our \textbf{validation set}. Aggregated \ac{DSC} is reported along with class-wise \ac{DSC} for GTVp and GTVn.}
\label{tab:result1}
\begin{tabular}{@{}l|l|l|l@{}}
\toprule
\bf{Method} & \bf{DSC} & \bf{GTVp} & \bf{GTVn} \\ \hline

SwinUNETR~\cite{hatamizadeh2022swin} & 0.7828  & 0.7121 & 0.6364 \\ \hline           
OARFocalFuseNet~\cite{srivastava2022efficient} & 0.7798 & 0.6898 & 0.6496 \\ \hline 
3D-MSF\cite{srivastava2022efficient} & 0.7951 & 0.7147 & 0.6706 \\ \hline

\end{tabular}
\end{table}

\section{Conclusion}
In this paper, we compared the performance of three multi-scale fusion methodologies for H\&N tumor segmentation to obtain accurate GTVn and GTVp. Our very recently proposed two multi-scale algorithms, originally designed for organ at-risk segmentation, were tuned for tumor segmentation from PET/CT scans obtained from the HECKTOR 2022 Challenge. We observed that under a supervised scenario, our proposed 3D-MSF and OARFocalFuseNet algorithms perform well on the HECKTOR 2022 H\&N segmentation challenge. We plan to extend the multi-scale fusion strategy by introducing domain adaptation or generalization strategies within the framework to further advance the performance on Hecktor 2022 H\&N Segmentation Challenge.

\section*{Acknowledgment}
 This project is supported by the NIH funding: R01-CA246704 and R01-CA240639. The computations in this paper were performed on equipment provided by the Experimental Infrastructure for Exploration of Exascale Computing (eX3), which is financially supported by the Research Council of Norway under contract 270053.

%
%
\bibliographystyle{splncs04}
\bibliography{references}
\end{document}